\documentclass[12pt]{article}

\usepackage{amssymb}
\usepackage{amsxtra}
\usepackage{amsmath}
\usepackage{amsfonts}
\usepackage{CJK}
\usepackage[titletoc]{appendix}
\usepackage{graphicx}
\usepackage{color}
\numberwithin{equation}{section}

\newtheorem{thm}{Theorem}[section]

\newtheorem{prop}[thm]{Proposition}
\newtheorem{hyp}[thm]{Hypothesis}

\newtheorem{de}[thm]{Definition}
\newtheorem{rem}[thm]{Remark}

\newcommand{\eqa}{\begin{eqnarray}}
\newcommand{\eeqa}{\end{eqnarray}}
\newcommand{\beq}{\begin{equation}}
\newcommand{\eeq}{\end{equation}}
\newcommand{\nn}{\nonumber}

\allowdisplaybreaks

\begin{document}
\title{ { A Riemann-Hilbert Approach to the Kundu-Eckhaus Equation on the half-Line }}

\author{{\footnotesize { Bei-bei Hu$^{a,b}$ , Tie-cheng Xia$^{a,}$\thanks{Corresponding author. E-mails: hubsquare@chzu.edu.cn; xiatc@shu.edu.cn; zhangningsdust@126.com} ,  Ning Zhang$^{a,c}$ }}\\
{\footnotesize { {\it$^{a}$Department of Mathematics, Shanghai University, Shanghai 200444, China}}}\\
{\footnotesize { \it $^{b}$School of Mathematics and Finance, Chuzhou University, Anhui 239000, China}}\\
{\footnotesize { \it $^{c}$Department of Basical Courses, Shandong University of Science and Technology, Taian 271019, China}}
}
\date{\small }\maketitle

\textbf{Abstract}: In this paper, we consider the initial-boundary value problem of the Kundu-Eckhaus equation on the half-line by using of the Fokas unified transform method. Assuming that the solution $u(x,t)$ exists, we show that it can be expressed in terms of the unique solution of a matrix Riemann-Hilbert problem formulated in the plane of the complex spectral parameter $\lambda$. Moreover, we also get that some spectral functions are not independent and satisfy the so-called global relation.

\section{Introduction}

It is well known that the inverse scattering transformation (IST) is an important method for
analysis initial value problems of integrable nonlinear equations in mathematical physics.
However, in many experimental environment and field situations, the
wave motion is initiated by what corresponds to the
imposition of boundary value conditions rather than initial value
conditions. This naturally leads to consider the formulation of
initial-boundary value (IBV) problems instead of a pure
initial value problems.

In 1997, Fokas announced a new unified approach based on the Riemann-Hilbert factorization problem to analysis the IBV problems for linear and nonlinear integrable PDEs \cite{Fokas1997,Fokas2002,Fokas2008}, we call that Fokas unified transform method. This method provides an important generalization of the IST formalism from initial value to IBV problems, and over the last 20 years, this method has been used to analyse boundary value problems for several of the most important integrable equations possessing $2\times2$ Lax pairs, such as the KdV, the nonlinear Schr\"{o}dinger(NLS), the sine-Gordon and the stationary axisymmetric Einstein equations and so on [4-15]. In 2012, Lenells first extended the Fokas unified transform method to the IBV problem for the $3\times3$ matrix Lax pair \cite{Lenells2012,Lenells2013}. After that, more and more researchers begin to pay attention to studying IBV problems for integrable evolution equations with higher order Lax pairs on the half-line or on the interval, the IBV problem for the many integrable equations with $3\times3$ or $4\times4$ Lax pairs are studied, such as, the Degasperis-Procesi equation \cite{Lenells2013,Monvel2013}, the Sasa-Satsuma equation \cite{Xu2013}, the three wave equation \cite{Xu2014}, the coupled NLS equation \cite{Geng2015}, the vector modified KdV equation \cite{Liu2016}, the Novikov equation \cite{Monvel2016}, the general coupled NLS equation\cite{Tian2017}. the integrable spin-1 Gross-Pitaevskii equations with a $4\times4$ Lax pair \cite{Yan2017}. These authors have also done some works about the IBV problem for integrable equations with $2\times2$ or $3\times3$ Lax pairs on the half-line \cite{Zhang2017,Hu2018,hu2017}. Just like the IST on the line, the Fokas unified transform method yields an expression for the solution of an IBV problem in terms of the solution of a Riemann-Hilbert problem (RHP). In particular, the asymptotic behaviour of the solution can be analysed in an effective way by using this Riemann-Hilbert problem and by employing the nonlinear version of the steepest descent method introduced by Deift and Zhou \cite{Deift1993}.

In this paper, we consider Kundu-Eckhaus(KE) equation as follows
\eqa iu_{t}+u_{xx}+2|u|^2u+4\beta^2|u|^4u-4i\beta(|u|^2)_{x}u=0, \beta\in R, \label{slisp}\eeqa
where $u(x,t)$ is the complex smooth envelop function, $t$ and $x$ are the temporal and spatial variables, respectively, and the last term represents the Raman effect, which accounts for the self-frequency shift of the waves in optics. The KE equation is an integrable equation and introduced independently during 1984 by Kundu \cite{Kundu1984} and later by Calogero and Eckhaus \cite{Calogero1987} from different perspective. The KE equation has been studied extensively on the integrability associated with explicit form of the Lax pair and Painlev\'{e} property \cite{calogero1987}, Hamiltonian structure \cite{Geng1992}, higher order extension \cite{Kundu2006}, Miura transformation \cite{Levi2009}, infinitely many conservation laws \cite{lu2013}, soliton given by the bilinear method \cite{Kakei1995} and other methods [37-40], rogue waves solutions given by the DT method [41-45].

Recently, Zhu etal. present a RHP formalism for the initial value problem of the KE equation on the line and using the Deift-Zhou nonlinear steepest descent method to analyzed the long-time asymptotic for the solutions of the KE equation in \cite{Zhu2018}. But the IBV problem of the KE equation on the half-line has not been studied. Therefore, we analyse the IBV problem
of the KE equation on the half-line by using the Fokas unified transform method in this paper. That it to say, in the quarter $(x; t)$-plane
$$\Omega=\{(x,t)|0<x<\infty,0<t<T\}$$
Assume that the solution $u(x; t)$ of the KE equation exists, and the initial-boundary
values data are defined as follows
\eqa\begin{array}{l}
$Initial values$: u_0(x)=u(x,t=0), 0<x<\infty;\\
$Boundary values$: g_0(t)=u(x=0,t),g_1(t)=u_{x}(x=0,t),0<t<T.
\end{array}\label{slisp}\eeqa
We will show that $u(x,t)$ can be expressed in terms of the unique solution of a matrix RHP formulated in the plane of the complex spectral parameter $\lambda$. The jump matrix has explicit $(x,t)$ dependence and is given in terms of the spectral functions $a(\lambda),b(\lambda)$ and $A(\lambda),B(\lambda)$, which obtained from the initial data $u_0(x)=u(x,0)$ and the boundary data $g_0(t)=u(0,t),g_1(t)=u_x(0,t)$, respectively. The problem has the jump across $\{Im\lambda^2=0\}$. The spectral functions are not independent, but related by a compatibility condition, the so-called global relation, which is an algebraic equation coupling $a(\lambda),b(\lambda)$ and $A(\lambda),B(\lambda)$.
Where a different RHP was formulated and a different representation of the solution of the KE equation was given, which are more convenient for studying the long-time asymptotic behavior for the solutions of the KE equation with the decay initial and boundary values lie in the Schwartz class.

Organization of this paper is as follows. In section 2, some summary results and the
basic RHP of the KE equation are given. In section 3, the spectral functions $a(\lambda), b(\lambda)$
and $A(\lambda), B(\lambda)$ are investigated and the RHP is presented. The
last section is devoted to conclusions and discussions.

\section{Spectral analysis}

We consider the following Lax pair of the KE equation \cite{Qiu2015}:
\eqa\left\{\begin{array}{l}
\phi_{x}=M\phi, \\
\phi_{t}=N\phi=(\lambda^{2}N_2+\lambda N_1+N_0)\phi,
\end{array}\right.\label{slisp}\eeqa
where
\eqa \begin{array}{l}
M=\left(\begin{array}{cc}
-i\lambda+i\beta|u|^2 & u \\
-\overline{u} & i\lambda-i\beta|u|^2
\end{array}\right),
N_2=\left(\begin{array}{cc}
-2i & 0 \\
 0 & 2i
\end{array}\right),
N_1=\left(\begin{array}{cc}
0 & 2u \\
-2\overline{u} & 0
\end{array}\right),\\
N_0=\left(\begin{array}{cc}
\beta(-u_x\overline{u}+u\overline{u}_x)+4i\beta^2|u|^4+i|u|^2  & iu_x+2\beta|u|^2u \\
 i\overline{u}_x-2\beta|u|^2u  & \beta(u_x\overline{u}-u\overline{u}_x)-4i\beta^2|u|^4-i|u|^2
\end{array}\right).
\end{array} \label{slisp}\eeqa
In the following section, we set $\beta=1$ and introducing
\eqa \begin{array}{l}
\sigma_{3}=\left(\begin{array}{cc}
1 & 0 \\
0 & -1
\end{array}\right),
\sigma_{1}=\left(\begin{array}{cc}
1& 0\\
0 & 0
\end{array}\right),
Q=\left(\begin{array}{cc}
0 & u \\
-\overline{u} & 0
\end{array}\right),
\end{array} \label{slisp}\eeqa
for the convenient of the analysis, where $\sigma_{3}$ denotes the third Pauli's matrix.

\subsection{Transformed Lax pair for the KE equation}

We can rewrite the Lax pair (2.1) in a matrix form
\eqa \left\{ \begin{array}{l}
\phi_{x}+i\lambda\sigma_{3}\phi=U_1\phi, \\
\phi_{t}+2i\lambda^2\sigma_{3}\phi=U_2\phi,
\end{array}\right. \label{slisp}\eeqa
where $U_1=-iQ^2\sigma_{3}+Q$ and $U_2=2\lambda Q+[Q_x,Q]+4iQ^4\sigma_{3}-iQ^2\sigma_{3}-iQ_x\sigma_{3}-2Q^3.$

Setting
\eqa \phi=\psi e^{-i(\lambda x+2\lambda^2t)\sigma_3},0<x<\infty, 0<t<T. \label{slisp}\eeqa
Then, we have the equivalent Lax pair
\eqa \left\{ \begin{array}{l}
\psi_x+i\lambda[\sigma_3,\psi]=U_1\psi,\\
\psi_t+2i\lambda^2[\sigma_3,\psi]=U_2\psi,
\end{array}\right.\label{slisp}\eeqa
which can be written as
\eqa d(e^{i(\lambda x+2\lambda^2t )\hat{\sigma}_3)}\psi(x,t;\lambda))=W_1(x,t;\lambda), 0<x<\infty, 0<t<T,\label{slisp}\eeqa
where
\eqa W_1(x,t;\lambda)=e^{i(\lambda x+2\lambda^2t )\hat{\sigma}_3}(U_1dx+U_2dt)\psi(x,t;\lambda),\eeqa
and $\hat{\sigma}_{3}$ denotes the matrix commutator with $\sigma_{3}$, $\hat{\sigma}_{3}A=[\sigma_{3},A]$, then $exp(\hat{\sigma}_{3})$ can be easily computed:  $e^{\hat{\sigma}_{3}}A=e^{\sigma_{3}}Ae^{-\sigma_{3}}$, where $A$ is a $2\times2$ matrix.

Next, we consider that a solution of Eq.(10) is of the form
\beq \psi(x,t;\lambda)=D_0+\frac{D_1}{\lambda}+O(\frac{1}{\lambda^2}), \lambda\rightarrow\infty.\eeq
Substituting Eq.(2.9) into the first equation of (2.6), and comparing the cofficient of $\lambda$ yields
\eqa  \begin{array}{l}
\lambda^1:i[\sigma_3,D_0]=0,\\
\lambda^0:D_{0x}+i[\sigma_3,D_1]=(-iQ^2\sigma_3+Q)D_0,\\
\lambda^{-1}:D_{1x}=(-iQ^2\sigma_3+Q)D_1.
\end{array}\label{slisp}\eeqa
We know that $D_0$ is a diagonal matrix form $O(\lambda^1)$, and set
$$D_0=\left(\begin{array}{cc}
D_0^{11} & 0 \\
0 & D_0^{22}
\end{array}\right).$$
From $O(\lambda^0)$ we have
\beq \begin{array}{cc}
D_1^{(o)}=\left(\begin{array}{cc}
0&  -\frac{i}{2}uD_0^{22} \\
-\frac{i}{2}\overline{u}D_0^{11} &0
 \end{array}\right),\,D_{0x}=i|u|^2\sigma_3D_0,\end{array}\eeq
where $D_1^{(o)}$ being the off-diagonal part of $D_1$.
On the other hand, substituting Eq.(2.9) into the second equation of (2.6), and comparing the cofficient of $\lambda$ yields
\eqa  \begin{array}{l}
\lambda^2:2i[\sigma_3,D_0]=0,\\
\lambda^1:2i[\sigma_3,D_1]=2QD_0,\\
\lambda^0:D_{0t}=([Q_x,Q]+4iQ^4\sigma_{3}-iQ^2\sigma_{3}-iQ_x\sigma_{3}-2Q^3)D_0+2QD_1,\\
\lambda^{-1}:D_{1t}=([Q_x,Q]+4iQ^4\sigma_{3}-iQ^2\sigma_{3}-iQ_x\sigma_{3}-2Q^3)D_1.
\end{array}\label{slisp}\eeqa
After a lengthy calculation, we get
\eqa D_{0t}=(-u_x\overline{u}+u\overline{u}_x+4i|u|^4)\sigma_3D_0.\label{slisp}\eeqa
The Eq.(2.1) have the following conservation law
$$i(u\overline{u})_t=(u_x\overline{u}-u\overline{u}_x+4iu^2\overline{u}^2)_x. $$
Then Eq.(2.11) and Eq.(2.13) for $D_0$ are consistent and are both satisfied if we define
\eqa D_0(x,t)=exp(i\int_{(x_0,t_0)}^{(x,t)}\Delta(x,t)\sigma_3), \label{slisp}\eeqa
where $\Delta$ is the closed real-valued one-form, and it is given by
$$\Delta(x,t)=\Delta_1dx+\Delta_2dt=u\overline{u}dx+(4u^2\overline{u}^2-i(u\overline{u}_x-u_x\overline{u})) dt, (x_0,t_0)\in D,$$
simultaneity, we let $(x_0,t_0)=(0,0)$ for the convenience of calculation.

Noting that the integral in Eq.(2.14) is independent of
the path of integration and the $\Delta$ is independent of $\lambda$, then we can introduce a new function $\mu(x,t;\lambda)$ as follows
\eqa \psi(x,t;\lambda)=e^{i\int_{(0,0)}^{(x,t)}\Delta\hat{\sigma}_3}\mu(x,t;\lambda)D_0(x,t),  0<x<\infty,  0<t<T. \label{slisp}\eeqa
Then the Lax pair of Eq.(2.7) can be written by
\eqa d(e^{i(\lambda x+2\lambda^2t)\hat{\sigma}_3}\mu(x,t;\lambda))=W_2(x,t;\lambda), \lambda\in\mathbb{C}, \label{slisp}\eeqa
where
\beq \begin{array}{cc}
W_2(x,t;\lambda)=e^{i(\lambda x+2\lambda^2t)\hat{\sigma}_3}V(x,t;\lambda)\mu(x,t;\lambda),\\
V(x,t;\lambda)=V_1(x,t;\lambda)dx+V_2(x,t;\lambda)dt=e^{-i\int_{(0,0)}^{(x,t)}\Delta\hat{\sigma}_3}(U_1dx+U_2dt-i\Delta\sigma_3).
\end{array}\eeq

By $U(x,t;\lambda)$ and $\Delta$, we can get
$$V_1(x,t;\lambda)=\left(\begin{array}{cc}
0 &  ue^{-2i\int_{(0,0)}^{(x,t)}\Delta} \\
-\overline{u}e^{2i\int_{(0,0)}^{(x,t)}\Delta} & 0
\end{array}\right),$$
and
$$ V_2(x,t;\lambda)=\left(\begin{array}{cc}
i|u|^2+2u\overline{u}_x-2u_x\overline{u} & (2\lambda u+iu_x+2u|u|^2)e^{-2i\int_{(0,0)}^{(x,t)}\Delta} \\
(-2\lambda \overline{u}+i\overline{u}_x-2u|u|^2)e^{2i\int_{(0,0)}^{(x,t)}\Delta} & -i|u|^2-2u\overline{u}_x+2u_x\overline{u}
\end{array}\right).$$
Then Eq.(2.16) can be written as
\eqa \left\{ \begin{array}{l}
\mu_x+i\lambda[\sigma_3,\mu]=V_1\mu,\\
\mu_t+2i\lambda^2[\sigma_3,\mu]=V_2\mu,
\end{array}\right.\label{slisp}\eeqa
where $0<x<\infty, 0<t<T, \lambda \in \mathbb{C}.$

\subsection{ Eigenfunctions and Some Relations}

Suppose that $u(x,t)$ sufficiently smooth in $D=\{0<x<\infty,0<t<T\}$, $\mu_j(x,t,\lambda) (j=1,2,3.)$ are the $2\times2$ matrix valued functions, and we defined the three eigenfunctions $\mu_j(x,t,\lambda) (j=1,2,3.)$ of (2.18) by the integral equations
\eqa \mu_j(x,t;\lambda)=I+\int_{(x_j,t_j)}^{(x,t)}e^{-i(\lambda x+2\lambda^2t)\hat{\sigma}_3}W_2(\xi,\tau,\lambda), 0<x<\infty, 0<t<T, \label{slisp}\eeqa
where the integral denotes a smooth curve from $(x_j,t_j)$ to $(x,t)$, and
$(x_1, t_1)=(0, T ), (x_2, t_2) = (0,0), (x_3, t_3) = (\infty, t)$ (see Figure 1).

\begin{figure}
\centering
\includegraphics[width=4.4in,height=1.4in]{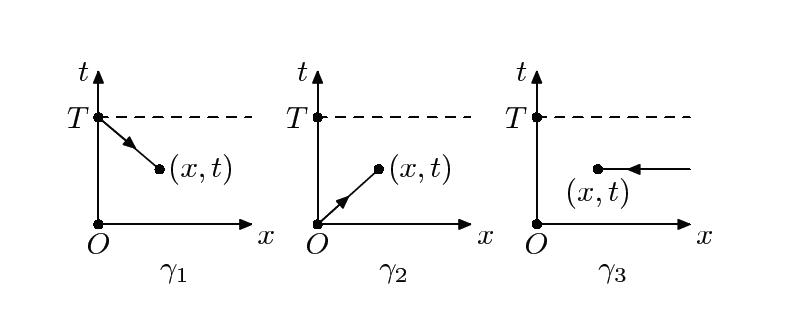}
\caption{The three contours $\gamma_1,\gamma_2,\gamma_3$ in the $(x,t)$-domaint}
\label{fig:graph}
\end{figure}

Since Eq.(2.19)are independent of the path of integration, we choose the specific contours depicted in Figure 1 yields
\eqa \left\{ \begin{array}{l}
\mu_1(x,t;\lambda)=I+\int_{0}^{x}e^{i\lambda (\xi-x)\hat{\sigma}_3}(V_1\mu_1)(\xi,t,\lambda)d\xi
-e^{-i\lambda x\hat{\sigma}_3}\int_{t}^{T}e^{2i\lambda^2(\tau-t)\hat{\sigma}_3}(V_2\mu_1)(0,\tau,\lambda)d\tau,\\
\mu_2(x,t;\lambda)=I+\int_{0}^{x}e^{i\lambda (\xi-x)\hat{\sigma}_3}(V_1\mu_2)(\xi,t,\lambda)d\xi
+e^{-i\lambda x\hat{\sigma}_3}\int_{0}^{t}e^{2i\lambda^2(\tau-t)\hat{\sigma}_3}(V_2\mu_2)(0,\tau,\lambda)d\tau,\\
\mu_3(x,t;\lambda)=I-\int_{x}^{\infty}e^{i\lambda (\xi-x)\hat{\sigma}_3}(V_1\mu_3)(\xi,t,\lambda)d\xi.
\end{array}\right.\label{slisp}\eeqa

The first column of the matrix Eq.(2.19) involves $e^{-2i[\lambda (\xi-x)+2\lambda^2(\tau-t)]}$,
and we have the following inequalities on the contours:
\eqa  \begin{array}{l}
\gamma_1: 0<\xi<x, t<\tau<T,\\
\gamma_2: 0<\xi<x, 0<\tau<t,\\
\gamma_3: 0<x<\infty.
\end{array}\label{slisp}\eeqa
So, these inequalities imply that the first column of the functions $\mu_j(x,t;\lambda)(j=1,2,3.)$
are bounded and analytical for $\lambda\in \mathbb{C}$ such that $\lambda$ belongs to
\eqa  \begin{array}{l}
\mu_1^{(1)}(x,t;\lambda): \{Im\lambda \geq 0\}\cap\{Im\lambda^2\leq 0\},\\
\mu_2^{(1)}(x,t;\lambda): \{Im\lambda \geq 0\}\cap\{Im\lambda^2\geq 0\},\\
\mu_3^{(1)}(x,t;\lambda): \{Im\lambda \leq 0\}.
\end{array}\label{slisp}\eeqa
In the same way, the second column of the matrix Eq.(2.19) involves the inverse of the above exponential, which is bounded and analytical in
\eqa  \begin{array}{l}
\mu_1^{(2)}(x,t;\lambda): \{Im\lambda \leq 0\}\cap\{Im\lambda^2\geq 0\},\\
\mu_2^{(2)}(x,t;\lambda): \{Im\lambda \leq 0\}\cap\{Im\lambda^2\leq 0\},\\
\mu_3^{(2)}(x,t;\lambda): \{Im\lambda \geq 0\}.
\end{array}\label{slisp}\eeqa
Then, we obtain
\eqa  \begin{array}{l}
\mu_1(x,t;\lambda)=(\mu_{1}^{D_2}(x,t;\lambda),\mu_{1}^{D_3}(x,t;\lambda)),\\
\mu_2(x,t;\lambda)=(\mu_{2}^{D_1}(x,t;\lambda),\mu_{2}^{D_4}(x,t;\lambda)),\\
\mu_3(x,t;\lambda)=(\mu_{3}^{D_3\cup D_4}(x,t;\lambda),\mu_{3}^{D_1\cup D_2}(x,t;\lambda)),
\end{array}\label{slisp}\eeqa
where $\mu_{j}^{D_i}$ denotes $\mu_j$ which is bounded and analytic for $\lambda \in D_i$, and $D_i=\{z\in \mathbb{C}|2k\pi+\frac{i-1}{2}\pi<Arg z<2k\pi +\frac{i}{2}\pi\}$, $j=1,2,3,  i=1,2,3,4,  k=0,\pm1,\pm2,\cdots$, $Arg z$ denotes the argument of the complex $\lambda$, (see Figure 2).

\begin{figure}
  \centering
  \includegraphics[width=3.0in,height=2.4in]{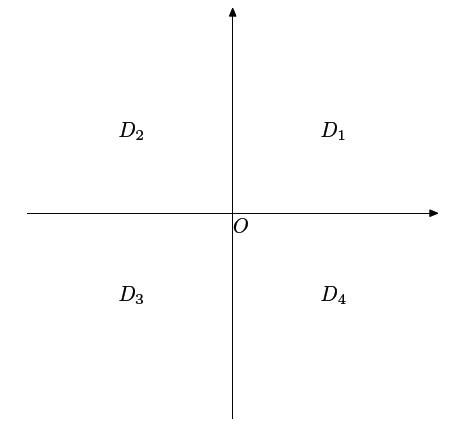}\\
  \caption{The sets $D_j,j=1,2,3,4$, which decompose the complex $\lambda-$plane}\label{fig:graph}
\end{figure}

More specifically,
\eqa \begin{array}{l}
\mu_1(0,t;\lambda)=(\mu_{1}^{D_2\cup D_4}(0,t;\lambda),\mu_{1}^{D_1\cup D_3}(0,t;\lambda)),\\
\mu_2(0,t;\lambda)=(\mu_{2}^{D_1\cup D_3}(0,t;\lambda),\mu_{2}^{D_2\cup D_4}(0,t;\lambda)),\\
\mu_1(x,T;\lambda)=(\mu_{1}^{D_1\cup D_2}(x,T;\lambda),\mu_{1}^{D_3\cup D_4}(x,T;\lambda)),\\
\mu_2(x,0;\lambda)=(\mu_{2}^{D_1\cup D_2}(x,0;\lambda),\mu_{2}^{D_3\cup D_4}(x,0;;\lambda)),\\
\mu_1(0,0;\lambda)=(\mu_{1}^{D_2\cup D_4}(0,0;\lambda),\mu_{1}^{D_1\cup D_3}(0,0;\lambda)),\\
\mu_2(0,T;\lambda)=(\mu_{2}^{D_1\cup D_3}(0,T;\lambda),\mu_{2}^{D_2\cup D_4}(0,T;\lambda)).
\end{array}\label{slisp}\eeqa

In order to deriving a RHP, we must to compute the jumps across the
boundaries of the $D_j$'s $(j=1,2,3,4.)$ at first. It turns out that the relevant jump matrices can be uniquely defined
in terms of two $2\times2$ matrices valued spectral functions $s(\lambda)$ and $S(\lambda)$ defined by
\eqa \left\{ \begin{array}{l}
\mu_3(x,t;\lambda)=\mu_2(x,t;\lambda)e^{-i(\lambda x+2\lambda^2t)\hat{\sigma}_3}s(\lambda),\\
\mu_1(x,t;\lambda)=\mu_2(x,t;\lambda)e^{-i(\lambda x+2\lambda^2t)\hat{\sigma}_3}S(\lambda).
\end{array}\right.\label{slisp}\eeqa
Calculating the first equation of (2.26) at $(x,t)=(0,0)$ and the second equation of (2.26) at $(x,t)=(0,T)$, we have
\eqa s(\lambda)=\mu_3(0,0;\lambda),S(\lambda)=(e^{2i\lambda^2 T\hat{\sigma}_3}\mu_2(0,T;\lambda))^{-1},\label{slisp}\eeqa
Then Eq.(2.26) and Eq.(2.27) implies that
\eqa \mu_1(x,t;\lambda)=\mu_3(x,t;\lambda)e^{-i(\lambda x+2\lambda^2t)\hat{\sigma}_3}(s(\lambda))^{-1}S(\lambda),\lambda\in(D_2\cup D_4,D_1\cup D_3), \label{slisp}\eeqa
which can be lead to the global relation.

As $\mu_1(0,t,\lambda)=\mathbb{I}$, when $(x,t)=(0,T)$, we can evaluate the following relationship which is the global relation as follows
\beq S^{-1}(\lambda)s(\lambda)=e^{2i\lambda^2T\hat\sigma_3}c(T,\lambda),\quad
\lambda\in(D_2\cup D_4,D_1\cup D_3),
\nn\eeq
where
\eqa c(T,\lambda)=\mu_3(0,t,\lambda)=I-\int_{0}^{\infty}e^{-i\lambda\xi\hat\sigma_3}(V_1\mu_3)(0,T,\lambda)d{\xi}.\nn\eeqa

Hence, the function $s(\lambda)$ and $S(\lambda)$ can be obtained from the evaluations at $x=0$ of the function
$\mu_3(x,0,\lambda)$ and at $t=T$ of the function $\mu_2(0,t,\lambda)$, respectively.  And these
functions about $\mu_j(x,t;\lambda)(j=1,2,3,)$ satisfy the linear integral equations
\eqa \begin{array}{l}
\mu_1(0,t;\lambda)=I-\int_{t}^{T}e^{2i\lambda^2(\tau-t)\hat{\sigma}_3}(V_2\mu_1)(0,\tau,\lambda)d\tau,\\
\mu_2(0,t;\lambda)=I+\int_{0}^{t}e^{2i\lambda^2(\tau-t)\hat{\sigma}_3}(V_2\mu_2)(0,\tau,\lambda)d\tau,\\
\mu_3(x,0;\lambda)=I-\int_{x}^{\infty}e^{i\lambda (\xi-x)\hat{\sigma}_3}(V_1\mu_3)(\xi,0,\lambda)d\xi,\\
\mu_2(x,0;\lambda)=I+\int_{0}^{x}e^{i\lambda (\xi-x)\hat{\sigma}_3}(V_1\mu_2)(\xi,0,\lambda)d\xi.
\end{array}\nn\eeqa
Let $u_0(x)=u(x,0)$, $g_0(t)=u(0,t)$, and $g_1(t)=u_x(0,t)$ be the initial and boundary values of $u(x,t)$, respectively.
$\overline{u}_0(x)=\overline{u}(x,0)$, $\overline{g}_0(t)=\overline{u}(0,t)$, and $\overline{g}_1(t)=\overline{u}_x(0,t)$ be the initial and boundary values of $\overline{u}(x,t)$, respectively. Then, we have
\eqa \begin{array}{c}
V_1(x,0;\lambda)=\left(\begin{array}{cc}
0 & u_0e^{-2i\int_{0}^{x}|u_0|^2d\xi} \\
-\bar{ u}_0e^{2i\int_{0}^{x}|u_0|^2d\xi}  & 0
\end{array}\right),\\
V_2(0,t;\lambda)=\left(\begin{array}{cc}
i|g_0|^2+2g_0\bar{g}_1-2g_1\bar{g}_0 & V_2^{(12)}(0,t;\lambda) \\
V_2^{(21)}(0,t;\lambda)  &  -i|g_0|^2-2g_0\bar{g}_1+2g_1\bar{g}_0
\end{array}\right),
\end{array}\label{slisp}\eeqa
where
$$V_2^{(12)}(0,t;\lambda)=(2\lambda g_0+ig_1+2|g_0|^2g_0)e^{-2i\int_{0}^{t}\Delta _0(0,\tau)d\tau},$$
$$V_2^{(21)}(0,t;\lambda)=(-2\lambda \bar{g}_0+i\bar{g}_1-2|g_0|^2\bar{g}_0)e^{2i\int_{0}^{t}\Delta _0(0,\tau)d\tau},$$
with $$\Delta _0(0,\tau)=4|g_0(\tau)|^2-i(g_0(\tau)\bar{g}_1(\tau)-g_1(\tau)\bar{g}_0(\tau)).$$
The analytic properties of $2\times2$ matrices $\mu_j(x,t;\lambda)(j=1,2,3.)$ that come from Eq.(2.19) are collected in the following proposition. We denote by $\mu_j^{(1)}(x,t;\lambda)$ and $\mu_j^{(2)}(x,t;\lambda)$ the first and second columns of $\mu_j(x,t;\lambda)$, respectively.

Setting
$$\mu_j(x,t;\lambda)=(\mu_j^{(1)}(x,t;\lambda),\mu_j^{(2)}(x,t;\lambda))
=\left(\begin{array}{cc}
   \mu_j^{11} & \mu_j^{12} \\
 \mu_j^{21} & \mu_j^{22} \\
\end{array}\right),j=1,2,3.$$

\begin{prop}
The matrices $\mu_j(x,t;\lambda)=(\mu_j^{(1)}(x,t;\lambda),\mu_j^{(2)}(x,t;\lambda)) (j=1,2,3.)$ have the following properties
\begin{itemize}
  \item $det\mu_1(x,t;\lambda)=det\mu_2(x,t;\lambda)=det\mu_3(x,t;\lambda)=1$;
  \item $\mu_1^{(1)}(x,t;\lambda)$ is analytic, and $\lim \limits_ {\lambda\rightarrow\infty}\mu_1^{(1)}(x,t;\lambda)=(1,0)^T, \lambda\in D_2$;
  \item $\mu_1^{(2)}(x,t;\lambda)$ is analytic, and $\lim \limits_ {\lambda\rightarrow\infty}\mu_1^{(2)}(x,t;\lambda)=(0,1)^T, \lambda\in D_3$;
  \item $\mu_2^{(1)}(x,t;\lambda)$ is analytic, and $\lim \limits_ {\lambda\rightarrow\infty}\mu_2^{(1)}(x,t;\lambda)=(1,0)^T, \lambda\in D_1$;
  \item $\mu_2^{(2)}(x,t;\lambda)$ is analytic, and $\lim \limits_ {\lambda\rightarrow\infty}\mu_2^{(2)}(x,t;\lambda)=(0,1)^T, \lambda\in D_4$;
  \item $\mu_3^{(1)}(x,t;\lambda)$ is analytic, and $\lim \limits_ {\lambda\rightarrow\infty}\mu_3^{(1)}(x,t;\lambda)=(1,0)^T, \lambda\in D_3\cup D_4$;
  \item $\mu_3^{(2)}(x,t;\lambda)$ is analytic, and $\lim \limits_ {\lambda\rightarrow\infty}\mu_3^{(2)}(x,t;\lambda)=(0,1)^T, \lambda\in D_1\cup D_2$.
\end{itemize}
\end{prop}

\begin{prop}
(Symmetries) The matrices
$$\mu_j(x,t;\lambda)=\left(
 \begin{array}{cc}
\mu_j^{11}(x,t;\lambda) & \mu_j^{12}(x,t;\lambda) \\
\mu_j^{21}(x,t;\lambda) & \mu_j^{22}(x,t;\lambda)
\end{array}\right)(j=1,2,3.),$$
have the following properties
\begin{itemize}
  \item \beq \mu_j^{11}(x,t;\lambda)=\overline{\mu_j^{22}(x,t;\bar{\lambda})},\, \mu_j^{12}(x,t;\lambda)=\overline{\mu_j^{21}(x,t;\bar{\lambda})}.\nn\eeq
  \item \eqa \begin{array}{l}
  \mu_j^{11}(x,t;-\lambda)=\mu_j^{11}(x,t;\lambda),\, \mu_j^{12}(x,t;-\lambda)=-\mu_j^{12}(x,t;\lambda), \\
  \mu_j^{21}(x,t;-\lambda)=-\mu_j^{21}(x,t;\lambda),\, \mu_j^{22}(x,t;-\lambda)=\mu_j^{22}(x,t;\lambda).\end{array}\nn\eeqa
\end{itemize}
\end{prop}

\begin{prop}
The spectral function $s(\lambda)$ and $S(\lambda)$ are defined in Eq.(2.26) and Eq.(2.27) mean that
\eqa \begin{array}{l}
s(\lambda)=I-\int_{0}^{\infty}e^{i\lambda \xi \hat{\sigma}_3}(V_1\mu_3)(\xi,0;\lambda)d\xi,\\
S(\lambda)=(I+\int_{0}^{T}e^{2i\lambda^2\tau\hat{\sigma}_3}(V_2\mu_2)(0,\tau;\lambda)d\tau)^{-1}.
\end{array}\label{slisp}\eeqa
\end{prop}

According to \textbf{Proposition 2.2}, we can define the following matrix functions $s(\lambda)$ and $S(\lambda)$,
\eqa
s(\lambda)=\left(\begin{array}{cc}
\overline{a(\bar{\lambda})} & b(\lambda)\\
\overline{b(\bar{\lambda})}& a(\lambda)
\end{array}\right),
S(\lambda)=\left(\begin{array}{cc}
\overline{A(\bar{\lambda})} & B(\lambda)\\
\overline{B(\bar{\lambda})}& A(\lambda)\\
\end{array}\right).
\label{slisp}\eeqa
By use of  Eq.(2.27) and Eq.(2.30), we can obtain
\begin{itemize}
  \item
  $$\begin{array}{cc}
   \left(\begin{array}{c} b(\lambda) \\ a(\lambda) \\ \end{array} \right)=\mu_3^{(2)}(0,0;\lambda)=\left(
   \begin{array}{c}
    \mu_3^{12}(0,0;\lambda) \\
    \mu_3^{22}(0,0;\lambda) \\
  \end{array}
 \right), \\
 \left(
  \begin{array}{c} -e^{-4i\lambda^2T}B(\lambda) \\ \overline{A(\bar{\lambda})} \\ \end{array} \right)=\mu_2^{(2)}(0,T;\lambda)=\left(
  \begin{array}{c}
    \mu_2^{12}(0,T;\lambda) \\
    \mu_2^{22}(0,T;\lambda) \\
  \end{array}
\right).
\end{array} $$
  \item $$\begin{array}{cc}
  \partial_x\mu_3^{(2)}(x,0;\lambda)+2i\lambda \sigma_1\mu_3^{(2)}(x,0;\lambda)=U_1(x,0;\lambda)\mu_3^{(2)}(x,0;\lambda),\,\lambda\in D_1\cup D_2,0<x<\infty,\\
  \partial_t\mu_2^{(2)}(0,t;\lambda)+4i\lambda^2\sigma_1\mu_2^{(2)}(0,t;\lambda)=U_2(0,t;\lambda)\mu_2^{(2)}(x,0;\lambda),\,\lambda\in D_2\cup D_4,0<t<T.
  \end{array}$$
 \item $$
  \begin{array}{cc}
  a(-\lambda)=a(\lambda),\,b(-\lambda)=-b(\lambda),\\
  A(-\lambda)=A(\lambda),\,B(-\lambda)=-B(\lambda).
  \end{array}$$
 \item $$ det s(\lambda)=det S(\lambda)=1.$$
 \item $$
  \begin{array}{cc}
  a(\lambda)=1+O(\frac{1}{\lambda}), b(\lambda)=O(\frac{1}{\lambda}), \lambda\rightarrow\infty, \lambda\in D_1\cup D_2,\\
A(\lambda)=1+O(\frac{1}{\lambda}), B(\lambda)=O(\frac{1}{\lambda}), \lambda\rightarrow\infty, \lambda\in D_1\cup D_3.
  \end{array}$$
 \end{itemize}

\subsection{The Basic Riemann-Hilbert Problem}

In order to discuss convenient we setting
\eqa \begin{array}{l}
\theta(\lambda)=\lambda x+2\lambda^2t,\\
\alpha(\lambda)=\overline{a(\bar{\lambda})}A(\lambda)-\overline{b(\bar{\lambda})}B(\lambda),\\
\beta(\lambda)=a(\lambda)B(\lambda)-b(\lambda)A(\lambda),\\
\delta(\lambda)=\overline{a(\bar{\lambda})}\beta(\lambda)+b(\lambda)\alpha(\lambda).
\end{array}\label{slisp}\eeqa
And $M(x,t;\lambda)$ defined by
\eqa \begin{array}{l}
M_{+}(x,t,\lambda)=(\mu_2^{D_1}(x,t,\lambda),\frac{\mu_3^{D_1\cup D_2}(x,t,\lambda)}{a(\lambda)}), \lambda\in D_1,\\
M_{-}(x,t,\lambda)=(\mu_1^{D_2}(x,t,\lambda),\frac{\mu_3^{D_1\cup D_2}(x,t,\lambda)}{\overline{\alpha(\bar{\lambda})}}), \lambda\in D_2,\\
M_{+}(x,t,\lambda)=(\frac{\mu_3^{D_3\cup D_4}(x,t,\lambda)}{\alpha(\lambda)},\mu_1^{D_3}(x,t,\lambda)), \lambda\in D_3,\\
M_{-}(x,t,\lambda)=(\frac{\mu_3^{D_3\cup D_4}(x,t,\lambda)}{\overline{a(\bar{\lambda})}},\mu_2^{D_4}(x,t,\lambda)), \lambda\in D_4.
\end{array}\label{slisp}\eeqa
These definitions mean that
\beq  det M(x,t;\lambda)=1, M(x,t;\lambda)=I+O(\frac{1}{\lambda}), \lambda\rightarrow\infty. \eeq

\begin{thm}
Assuming that $u(x,t;\lambda)$ is a  sufficiently smooth function, $\mu_j(x,t,\lambda),(j=1,2,3)$ are defined in Eq.(2.19), and $M(x,t;\lambda)$ is defined in Eq.(2.33), then $M(x,t;\lambda)$ satisfies the jump condition on $\bar D_n\cap\bar D_m(n,m=1,2,3,4)$
\eqa M_{+}(x,t,\lambda)=M_{-}(x,t,\lambda)J(x,t,\lambda), \lambda \in \bar D_n\cap\bar D_m,\,n,m=1,2,3,4; n\neq m,\label{slisp}\eeqa
where
\eqa
J(x,t,\lambda)=\left\{ \begin{array}{l}
J_1(x,t,\lambda), \hbox{$Arg \lambda =0$,} \\
J_2(x,t,\lambda), \hbox{$Arg \lambda =\frac{\pi}{2}$,} \\
J_3(x,t,\lambda), \hbox{$Arg \lambda =\pi$,} \\
J_4(x,t,\lambda), \hbox{$Arg \lambda =\frac{3\pi}{2}$,}
\end{array}\right.\label{slisp}\eeqa
and
\eqa \begin{array}{l}
J_1(x,t,\lambda)=\left(\begin{array}{cc}
1 & \frac{b(\lambda)}{a(\lambda)}e^{-2i\theta(\lambda)} \\
-\frac{\overline{b(\bar{\lambda})}}{\overline{a(\bar{\lambda})}}e^{2i\theta(\lambda)} & \frac{1}{a(\lambda)\overline{a(\bar{\lambda})}}
\end{array}\right),\\
J_2(x,t,\lambda)=\left(\begin{array}{cc}
 \frac{a(\lambda)}{\overline{\alpha(\bar{\lambda})}} & 0 \\
 -\overline{\delta(\bar{\lambda})} e^{2i\theta(\lambda)}& \frac{\overline{\alpha(\bar{\lambda})}}{a(\lambda)}
 \end{array}\right),\\
J_3(x,t,\lambda)=\left(\begin{array}{cc}
\frac{1}{\alpha(\lambda)\overline{\alpha(\bar{\lambda})}} & \frac{\beta(\lambda)}{\overline{\alpha(\bar{\lambda})}}e^{-2i\theta(\lambda)} \\
 -\frac{\overline{\beta(\bar{\lambda})}}{\alpha(\lambda)}e^{2i\theta(\lambda)} & 1
\end{array}\right),\\
J_4(x,t,\lambda)=\left(\begin{array}{cc}
\frac{\overline{a(\bar{\lambda})}}{\alpha(\lambda)} & \delta(\lambda)e^{-2i\theta(\lambda)} \\
0 & \frac{\alpha(\lambda)}{\overline{a(\bar{\lambda})}}
 \end{array} \right).
\end{array}\nn\eeqa
\end{thm}

\textbf{Proof} We can complete the proof as \textbf{Proposition 2.2}'s idea in \cite{Fokas2005}. We can write Eq.(2.26) and Eq.(2.31) in the following form
\eqa \left\{ \begin{array}{l}
\overline{a(\bar{\lambda})}\mu_2^{D_1}+\overline{b(\bar{\lambda})}e^{2i\theta(\lambda)}\mu_2^{D_4}=\mu_3^{D_3\cup D_4},  \\
b(\lambda)e^{-2i\theta(\lambda)}\mu_2^{D_1}+a(\lambda)\mu_2^{D_4}=\mu_3^{D_1\cup D_2},
\end{array}\right.\label{slisp}\eeqa

\eqa \left\{ \begin{array}{l}
\overline{A(\bar{\lambda})}\mu_2^{D_1}+\overline{B(\bar{\lambda})}e^{2i\theta(\lambda)}\mu_2^{D_4}=\mu_1^{D_2},  \\
B(\lambda)e^{-2i\theta(\lambda)}\mu_2^{D_1}+A(\lambda)\mu_2^{D_4}=\mu_1^{D_3},
\end{array}\right.\label{slisp}\eeqa

\eqa \left\{ \begin{array}{l}
\overline{\alpha(\bar{\lambda})}\mu_3^{D_3\cup D_4}+\overline{\beta(\bar{\lambda})}e^{2i\theta(\lambda)}\mu_2^{D_1\cup D_2}=\mu_1^{D_2},  \\
\beta(\lambda)e^{-2i\theta(\lambda)}\mu_3^{D_3\cup D_4}+\alpha(\lambda)\mu_2^{D_1\cup D_2}=\mu_1^{D_3}.
\end{array}\right.\label{slisp}\eeqa

Using the Eq.(2.37), Eq.(2.38) and Eq.(2.39), it is not difficult to derive that the jump matrices $J_i(x,t;\lambda)(i=1,2,3,4.)$ satisfy
\eqa \begin{array}{l}
(\mu_2^{D_1}(x,t,\lambda),\frac{\mu_3^{D_1\cup D_2}(x,t,\lambda)}{a(\lambda)})=(\frac{\mu_3^{D_3\cup D_4}(x,t,\lambda)}{\overline{a(\bar{\lambda})}},\mu_2^{D_4}(x,t,\lambda))J_1(x,t;\lambda),\\
(\mu_2^{D_1}(x,t,\lambda),\frac{\mu_3^{D_1\cup D_2}(x,t,\lambda)}{a(\lambda)})=(\mu_1^{D_2}(x,t,\lambda),\frac{\mu_3^{D_1\cup D_2}(x,t,\lambda)}{\overline{\alpha(\bar{\lambda})}})J_2(x,t;\lambda),\\
(\frac{\mu_3^{D_3\cup D_4}(x,t,\lambda)}{\alpha(\lambda)},\mu_1^{D_3}(x,t,\lambda))=(\mu_1^{D_2}(x,t,\lambda),\frac{\mu_3^{D_1\cup D_2}(x,t,\lambda)}{\overline{\alpha(\bar{\lambda})}})J_3(x,t;\lambda),\\
(\frac{\mu_3^{D_3\cup D_4}(x,t,\lambda)}{\alpha(\lambda)},\mu_1^{D_3}(x,t,\lambda))=(\frac{\mu_3^{D_3\cup D_4}(x,t,\lambda)}{\overline{a(\bar{\lambda})}},\mu_2^{D_4}(x,t,\lambda))J_4(x,t;\lambda).
\end{array}\label{slisp}\eeqa

The matrix $M(x,t;\lambda)$ of this RHP is a sectionally meromorphic function
of $\lambda$ in $\bar D_n\cap\bar D_m(n,m=1,2,3,4)$. The possible poles of $M(x,t;\lambda)$ are generated by the zeros of $a(\lambda), \alpha(\lambda)$ and by the complex
conjugates of these zeros. Since $a(\lambda)$, $\alpha(\lambda)$ are even functions, this means each zero $\lambda_j$ of $a(\lambda)$ is accompanied by another zero at $-\lambda_j$.
Similarly, each zero $\lambda_j$ of $\alpha(\lambda)$ is accompanied by a zero at $-\lambda_j$. In particular, both $a(\lambda)$ and
$\alpha(\lambda)$ have even number of zeros.\\

\begin{hyp}
Suppose that
\begin{itemize}
  \item $a(\lambda)$ has $2n$ simple zeros $\{\varepsilon_j\}_{j=1}^{2n}$, $2n=2n_1+2n_2$, such that $\varepsilon_j$( $j=1,2,\cdots,2n_1$) lie in $D_1$, and $\bar{\varepsilon}_j$($j=1,2,\cdots,2n_2$) lie in $D_4$.
  \item $\alpha(\lambda)$ has $2N$ simple zeros $\{\gamma_j\}_{j=1}^{2N}$($2N=2N_1+2N_2$), such that $\gamma_j$($j=1,2,\cdots,2N_1$), lie in $D_3$, and $\bar{\gamma}_j$($j=1,2,\cdots,2N_2$), lie in $D_2$.
  \item None of the zeros of $\alpha(\lambda)$ coincides with any of the zeros of $a(\lambda)$.
\end{itemize}
\end{hyp}

\begin{prop}
(The residue conditions) The residues of the function $M(x,t;\lambda)$ at the corresponding poles can be computed using Eq.(2.26) and Eq.(2.28). Using the notation $[M(x,t;\lambda)]_1$ for the first column and $[M(x,t;\lambda)]_2$ for the second column of the solution $M(x,t;\lambda)$ of the RHP, respectively. Denote $\dot{a}(\lambda)=\frac{da}{d\lambda}$, then the following residue conditions is established:
\begin{description}
  \item[(i)]  Res$ \{[M(x,t;\lambda)]_{2} ,  \varepsilon_j\}=\frac{e^{-2i\theta(\varepsilon_j)}b(\varepsilon_j)}{\dot{a}(\varepsilon_j)}[M(x,t;\varepsilon_j)]_{1}, j=1,2,\cdots,2n_1$.
  \item[(ii)] Res $\{[M(x,t;\lambda)]_{1} ,  \bar{\varepsilon}_j \}$=$\frac{e^{2i\theta(\bar{\varepsilon_j})}\overline{b(\bar{\varepsilon}_j)}}{\overline{\dot{a}(\bar{\varepsilon}_j)}}[M(x,t;\bar{\varepsilon}_j)]_{2}$, $j=1,2,\cdots,2n_2$.
  \item[(iii)] Res $\{[M(x,t;\lambda)]_{1} ,  \gamma_j \}$=$\frac{e^{2i\theta(\gamma_j)}}{\dot{\alpha}(\gamma_j)\beta(\gamma_j)}[M(x,t;\gamma_j)]_{2}$, $j=1,2,\cdots,2N_1$.
  \item[(iv)] Res $\{[M(x,t;\lambda)]_{2} , \ \bar{\gamma}_j \}$=$\frac{e^{-2i\theta(\bar{\gamma}_j)}}{\overline{\dot{\alpha}(\bar{\gamma}_j)}\overline{\beta(\bar{\gamma_j})}}[M(x,t;\bar{\gamma_j})]_{1}$, $j=1,2,\cdots,2N_2$.
\end{description}
\end{prop}

\textbf{Proof} According to the idea in ref.\cite{Fokas2005}, we only need to prove \textbf{(i)}, and another three relations also have similar proof.

Consider $M(x,t;\lambda)=(\mu_2^{D_1},\frac{\mu_3^{D_1\cup D_2}}{a(\lambda)})$, the simple zeros $\varepsilon_j$ ($j=1,2,\cdots,2n_1$) of $a(\lambda)$ are the simple poles of $\frac{\mu_3^{D_1\cup D_2}}{a(\lambda)}$. Then we have
\beq Res\{\frac{\mu_3^{D_1\cup D_2}(x,t;\lambda)}{a(\lambda)},\varepsilon_j\}=\lim_{\lambda\rightarrow\varepsilon_j}(\lambda-\varepsilon_j)\frac{\mu_3^{D_1\cup D_2}(x,t;\lambda)}{a(\lambda)}=\frac{\mu_3^{D_1\cup D_2}(x,t;\varepsilon_j)}{\dot{a}(\varepsilon_j)}.\nn\eeq
Taking $\lambda=\varepsilon_j$ into the second equation of Eq.(2.37) yields
\beq \mu_3^{D_1\cup D_2}(x,t;\varepsilon_j)=b(\varepsilon_j)e^{-2i\theta(\varepsilon_j)}\mu_2^{D_1}(x,t;\varepsilon_j).\nn\eeq
Furthermore,
\beq Res\{\frac{\mu_3^{D_1\cup D_2}(x,t;\lambda)}{a(\lambda)},\varepsilon_j\}=\frac{b(\varepsilon_j)e^{-2i\theta(\varepsilon_j)}}{\dot{a}(\varepsilon_j)}\mu_2^{D_1}(x,t;\varepsilon_j).\nn\eeq
It is equivalent to \textbf{Proposition 2.6(i)}.

\subsection{ The Inverse Problem}

It is not difficult to see that the jump condition can be written as
\eqa
M_+(x,t;\lambda)-M_{-}(x,t;\lambda)=M_-\tilde{J}(x,t;\lambda),\eeqa
where $\tilde{J}(x,t;\lambda)=J(x,t;\lambda)-I$. The asymptotic conditions of Eq.(2.20) and the \textbf{Proposition 2.1} mean that
\eqa
M(x,t;\lambda)=I+\frac{\overline{M}(x,t;\lambda)}{\lambda}+O(\frac{1}{\lambda^2}),\,\lambda\rightarrow\infty,\,\lambda\in \mathbb{C} \setminus \Gamma,
\eeqa
where $\Gamma=\{\lambda^2=\mathbb{R}\}$.

Using Eq.(2.41) and Eq.(2.42) not difficult to find that
\eqa
M(x,t;\lambda)=I+\frac{1}{2\pi i}\int_{\Gamma}\frac{M_{+}(x,t;\lambda')\tilde{J}(x,t;\lambda')}{\lambda-\lambda'}d\lambda',\,\lambda\in \mathbb{C} \setminus \Gamma,
\nn\eeqa
then
\eqa
\overline{M}(x,t;\lambda)=-\frac{1}{2\pi i}\int_{\Gamma}M_{+}(x,t;\lambda')\tilde{J}(x,t;\lambda')d\lambda',
\nn\eeqa
and using Eq.(2.42) in the first ODE of the Lax pair Eq.(2.18) yields
\eqa
 -[\sigma_3,\overline{M}(x,t;\lambda)]=[u_x(x,t)-iu_t(x,t)]\sigma_1,\nn\eeqa
\eqa u_x(x,t)-iu_t(x,t)=2(\overline{M(x,t;\lambda)})_{21}=2\lim_{\lambda\rightarrow\infty}(\lambda M(x,t;\lambda))_{21},\nn\eeqa
where $\sigma_1$ and $\sigma_3$ being the usual Pauli matrices defined in (2.3).

The inverse problem involves reconstructing the potential $u(x,t)$ from the spectral functions
$\mu_j (j=1,2,3)$. That means we will reconstruct the potential $u(x,t)$. We
show in \textbf{Section 2.2} that
$$D_1^{(o)}=\left(\begin{array}{cc}
 0&  -\frac{i}{2}uD_0^{22} \\
 -\frac{i}{2}\overline{u}D_0^{11} &0
\end{array} \right),$$
when
$\psi(x,t;\lambda)=D_0+\frac{D_1}{\lambda}+O(\frac{1}{\lambda^2})\quad(\lambda\rightarrow\infty)$
is a solution of Eq.(2.7). This implies that
\eqa u(x,t)=2im(x,t)e^{2i\int_{(0,0)}^{(x,t)}\Delta},\label{slisp}\eeqa
where
$$\mu(x,t;\lambda)=I+\frac{m^{(1)}(x,t;\lambda)}{\lambda}+O(\frac{1}{\lambda^2}), (\lambda\rightarrow\infty),$$
is the corresponding solution of Eq.(2.16) related to $\psi(x,t;\lambda)$ via Eq.(2.15), and we write $m(x,t)$ for $m^{(1)}_{12}(x,t)$. From Eq.(2.43) and its complex conjugate, we obtain
$$u\overline{u}=4|m|^2,\quad u\overline{u}_x-u_x\overline{u}=4(\bar{m}_xm-m_x\bar{m})-32i|m|^4.$$
Thus, we are able to express the one-form $\Delta$ defined in Eq.(2.14) in terms of $m(x,t;\lambda)$ as
\eqa
\Delta=4|m|^2dx+(48|m|^4-4i(\bar{m}_xm-m_x\bar{m}))dt.
\label{slisp}\eeqa
Then we can solve the inverse problem as follows
\begin{description}
  \item[(i)] Use any one of the three spectral functions $\mu_j$ to compute $m(x,t)$ according to
$$m(x,t)=\lim_{\lambda\rightarrow\infty}(\lambda\mu_j(x,t;\lambda))_{12}.$$
  \item[(ii)] Determine $\Delta(x,t)$ from Eq.(2.44).
  \item[(iii)] Finally, $u(x,t)$ is given by Eq.(2.43).
\end{description}

\section{The Spectral Functions and the Riemann-Hilbert Problem}

\subsection{The Spectral Functions}

\begin{de}
 (The spectral functions $a(\lambda)$ and $b(\lambda)$) Given the smooth function $u_{0}(x)=u(x,0)$,
we define the map
$$\mathbb{S}: \{u_0(x)\}\rightarrow \{a(\lambda),b(\lambda) \},$$
by
$$
\left(\begin{array}{c}
    b(\lambda) \\
    a(\lambda)
  \end{array}\right)
=\mu_3^{(2)}(x,0;\lambda)=\left(\begin{array}{c}
    \mu_3^{12}(x,0;\lambda) \\
    \mu_3^{22}(x,0;\lambda) \\
  \end{array}\right),  Im\lambda \geq 0,
$$
where $\mu_3(x,0;\lambda)$ is the unique solution of the Volterra linear integral equation
$$\mu_3(x,0;\lambda)=I-\int_{x}^{\infty }e^{i\lambda (\xi-x)\hat{\sigma}_3}(V_1\mu_3)(\xi,0;\lambda)d\xi,$$
and $V_1(x, 0;\lambda)$ is given in terms of $u(x,0;\lambda)$ in Eq.(2.29).
\end{de}

\begin{prop}
The spectral functions $a(\lambda)$ and $b(\lambda)$ have the
following properties
\begin{description}
  \item[(i)] $a(\lambda)$ and $b(\lambda)$ are analytic and bounded for $Im\lambda < 0$;
  \item[(ii)]$a(\lambda)=1+O(\frac{1}{\lambda}),b(\lambda)=O(\frac{1}{\lambda})$ as $\lambda\rightarrow\infty$, $Im\lambda \geq 0$;
  \item[(iii)] $a(\lambda)\overline{a(\bar{\lambda})}-b(\lambda)\overline{b(\bar{\lambda})}=1$, $\lambda \in \mathbb{R}$;
  \item[(iv)] $a(-\lambda)=a(\lambda),b(-\lambda)=-b(\lambda)$, $Im\lambda \geq 0$;
  \item[(v)] $\mathbb{Q}=\mathbb{S}^{-1}: \{a(\lambda),b(\lambda) \}\rightarrow \{u_0(x)\}$, the inverse map $\mathbb{S}$ of $\mathbb{Q}$ is defined by
$u_0(x)=2im(x)$,$m(x)=\lim_{\lambda\rightarrow\infty}(\lambda M^{(x)}(x,\lambda))_{12}$
where, $M^{(x)}(x,\lambda)$ is the unique solution of the following RHP, (see \textbf{Remark 3.3}).
\end{description}
\end{prop}

\begin{rem}
The \textbf{Definition 3.1} gives rise to the map
$$\mathbb{S}: \{u_0(x)\}\rightarrow \{a(\lambda),b(\lambda) \},$$
the inverse of this map
$$\mathbb{Q}: \{a(\lambda),b(\lambda) \}\rightarrow \{u_0(x)\},$$
can be defined as follows
\eqa
u_0(x)=2im(x)e^{2i\int_{0}^{x}\Delta_1(\xi)d\xi},\,
m(x)=\lim_{\lambda\rightarrow\infty}(\lambda M^{(x)}(x,\lambda))_{12},
\eeqa
where $\Delta_1(\xi)$ defined by Eq.(2.14) and $M^{(x)}(x,\lambda)$ is the unique solution of the following RHP.
\end{rem}

\begin{itemize}
  \item $M^{(x)}(x,\lambda)=\left\{ \begin{array}{l}
  M_{-}^{(x)} (x,\lambda),  \lambda\in D_1\cup D_2 \\
  M_{+}^{(x)} (x,\lambda),  \lambda\in D_3\cup D_4
\end{array}\right.$ is a sectionally meromorphic function.
  \item $M_{+}^{(x)} (x,\lambda)=M_{-}^{(x)} (x,\lambda)(J^{(x)} (x,\lambda))^{-1}$, $\lambda\in \mathbb{R}$, and
\eqa J^{(x)} (x,\lambda)=\left(\begin{array}{cc}
1 & \frac{b(\lambda)}{a(\lambda)}e^{-2i\lambda x} \\
-\frac{\overline{b(\bar{\lambda})}}{\overline{a(\bar{\lambda})}}e^{2i\lambda x} & \frac{1}{a(\lambda)\overline{a(\bar{\lambda})}}
\end{array}\right). \label{slisp}\eeqa
  \item $M^{(x)} (x,\lambda)=I+O(\frac{1}{\lambda}), \lambda\rightarrow\infty.$
  \item $a(\lambda)$ has $2n$ simple zeros $\{\varepsilon_j\}_{1}^{2n}$, $2n=2n_1+2n_2$, such that, $\varepsilon_j$($j=1,2,\cdots,2n_1$) lie in $D_3\cup D_4$, $\bar{\varepsilon}_j$($j=1,2,\cdots,2n_2$) lie in $D_1\cup D_2$.
  \item The first column of $M_{-}^{(x)}(x,\lambda)$ has simple poles at $\lambda=\bar{\varepsilon}_j$, $j=1,2,\cdots,2n_2$. The second column of $M_{+}^{(x)}(x,\lambda)$ has simple poles at $\lambda=\varepsilon_j$, $j=1,2,\cdots,2n_1$.
The associated residues are given by
\eqa  Res \{[M^{(x)}(x,\lambda)]_{1} ,  \bar{\varepsilon}_j \}=\frac{e^{2i\bar{\varepsilon_j}x}\overline{b(\bar{\varepsilon}_j)}}{\overline{\dot{a}(\bar{\varepsilon}_j)}}[M^{(x)}(x,\bar{\varepsilon}_j)]_{2}, j=1,2,\cdots,2n_2,\label{slisp}\eeqa
\eqa  Res \{[M^{(x)}(x,\lambda)]_{2} ,  \varepsilon_j \}=\frac{e^{-2i\varepsilon_j x}b(\varepsilon_j)}{\dot{a}(\varepsilon_j)}[M^{(x)}(x,\varepsilon_j)]_{1}, j=1,2,\cdots,2n_1.\label{slisp}\eeqa
\end{itemize}

\begin{de}
(The spectral functions $A(\lambda)$ and $B(\lambda)$) Let $g_{0}(t)$, $g_{1}(t)$ be smooth functions,
we define the map
$$\mathbb{\tilde{S}}: \{g_0(t),g_1(t)\}\rightarrow \{A(\lambda),B(\lambda) \},$$
by
$$\left(\begin{array}{c}
    B(\lambda) \\
    A(\lambda)
  \end{array}\right)
=\mu_1^{(2)}(0,t,\lambda)=\left(\begin{array}{c}
    \mu_1^{12}(0,t,\lambda) \\
    \mu_1^{22}(0,t,\lambda)
  \end{array}\right),  Im\lambda\leq 0,$$
where $\mu_1(0,t,\lambda)$ is the unique solution of the Volterra linear integral equation
$$\mu_1(0,t,\lambda)=I-\int_{t}^{T}e^{2i\lambda^2(\tau-T)\hat{\sigma}_3}(V_2\mu_1)(0,\tau,\lambda)d\tau,$$
and $V_2(0, T;\lambda)$ is given in Eq.(2.29).
\end{de}

\begin{prop}
The spectral functions $A(\lambda)$ and $B(\lambda)$ have the following properties
\begin{description}
  \item[(i)] $A(\lambda)$ and $B(\lambda)$ are analytic for $Im\lambda^2>0$ and continuous and bounded for $Im\lambda^2\leq 0$;
  \item[(ii)]$A(\lambda)=1+O(\frac{1}{\lambda}),B(\lambda)=O(\frac{1}{\lambda})$ as $\lambda\rightarrow\infty$, $Im\lambda^2 \leq 0$;
  \item[(iii)] $A(\lambda)\overline{A(\bar{\lambda})}-B(\lambda)\overline{B(\bar{\lambda})}=1$, $\lambda^2 \in \mathbb{R}$;
\item[(iv)] $A(-\lambda)=A(\lambda),B(-\lambda)=-B(\lambda)$, $Im\lambda^2 \leq 0$;
\item[(v)] $\mathbb{\tilde{Q}}=\mathbb{\tilde{S}}^{-1}: \{A(\lambda),B(\lambda) \}\rightarrow \{g_0(t),g_1(t)\}$, the inverse map  $\mathbb{\tilde{S}}$ of $\mathbb{\tilde{Q}}$ is defined by
\eqa g_0(t)=2im_{12}^{(1)}(t)e^{2i\int_{0}^{t}\Delta_2(\tau)d\tau},\label{slisp}\eeqa
\eqa g_1(t)=(4m_{12}^{(1)}(t)-2|g_0(t)|^2)e^{2i\int_{0}^{t}\Delta_2(\tau)d\tau}+ig_0(t)(4m_{12}^{(1)}(t)+|g_0(t)|^2),\label{slisp}\eeqa
where $\Delta_2(\tau)$ is given in Eq.(2.14) and the function $m^{(1)}(t)$ determined by the asymptotic expansion
$$M^{(t)}(t,\lambda)=I+\frac{m^{(1)}(t,\lambda)}{\lambda}+O(\frac{1}{\lambda^2}),\,\lambda\rightarrow\infty,$$
where $M^{(t)}(t,\lambda)$ is the unique solution of the following RHP (see \textbf{Remark 3.6}).
\end{description}
\end{prop}

\begin{rem} Let
\eqa \begin{array}{l}
M_{+}^{(t)}(t,\lambda)=(\frac{\mu_{2}^{D_1\cup D_3}(t,\lambda)}{A(\lambda)},\mu_{1}^{D_1\cup D_3}(t,\lambda)),\,Im\lambda^2\geq 0,\\
M_{-}^{(t)}(t,\lambda)=(\mu_{1}^{D_2\cup D_4}(t,\lambda),\frac{\mu_{2}^{D_2\cup D_4}(t,\lambda)}{\overline{A(\bar{\lambda})}}),\,Im\lambda^2\leq 0,
\end{array}\label{slisp}\eeqa
$M^{(t)}(t,\lambda)$ is the unique solution of the following RHP.
\end{rem}

\begin{itemize}
  \item $M^{(t)}(t,\lambda)=\left\{
                             \begin{array}{ll}
                              M_{-}^{(t)} (t,\lambda), & \lambda\in D_1\cup D_3 \\
                              M_{+}^{(t)} (t,\lambda), & \lambda\in D_2\cup D_4
                             \end{array}
                           \right.$ is a sectionally meromorphic function.
  \item $M_{+}^{(t)} (t,\lambda)=M_{-}^{(t)} (t,\lambda)J^{(t)} (t,\lambda)$, $\lambda^2 \in \mathbb{R}$, and
\eqa
J^{(t)} (t,\lambda)=\left(
                     \begin{array}{cc}
                       \frac{1}{A(\lambda)\overline{A(\bar{\lambda})}} & \frac{B(\lambda)}{\overline{A(\bar{\lambda})}}e^{-4i\lambda^2 t} \\
                      -\frac{\overline{B(\bar{\lambda})}}{A(\lambda)}e^{4i\lambda^2 t} & 1 \\
                     \end{array}
                   \right).
\label{slisp}\eeqa
  \item $M^{(t)} (T,\lambda)=I+O(\frac{1}{\lambda}), \lambda\rightarrow\infty.$
  \item $A(\lambda)$ has $2k$ simple zeros $\{\zeta_j\}_{1}^{2k}$, $2k=2k_1+2k_2$, such that, $\zeta_j$($j=1,2,\cdots,2k_1$) lie in $D_1\cup D_3$, $\bar{\zeta}_j$($j=2k_1+1,2k_1+2,\cdots,2k$) lie in $D_2\cup D_4$.
  \item The first column of $M_{+}^{(t)}(t,\lambda)$ has simple poles at $\lambda=\zeta_j$, $j=1,2,\cdots,2k_1$. The second column of $M_{-}^{(t)}(t,\lambda)$ has simple poles at $\lambda=\bar{\zeta}_j$, $j=1,2,\cdots,2k_2$.\\
The associated residues are given by
\eqa Res \{[M^{(t)}(t,\lambda)]_{1} ,  \zeta_j \}=\frac{e^{4i\zeta_j^2t}}{\dot{A}(\zeta_j)B(\zeta_j)}[M^{(t)}(t,\zeta_j)]_{2}, j=1,2,\cdots,2k_1,\label{slisp}\eeqa
\eqa  Res \{[M^{(t)}(t,\lambda)]_{2} ,  \bar{\zeta}_j \}=\frac{e^{-4i\zeta_j^2t}}{\overline{\dot{A}(\bar{\zeta}_j)}\overline{B(\bar{\zeta}_j)}}[M^{(t)}(t,\bar{\zeta}_j)]_{1}, j=1,2,\cdots,2k_2.\eeqa
\end{itemize}

\begin{de}
(The spectral functions $\alpha(\lambda)$ and $\beta(\lambda)$)  Given the spectral functions
$$\alpha(\lambda)=\overline{a(\bar{\lambda})}A(\lambda)-\overline{b(\bar{\lambda})}B(\lambda),\,\beta(\lambda)=a(\lambda)B(\lambda)-b(\lambda)A(\lambda),$$
and the smooth functions $h_T(x)=u(x,T)$.
We define the map
$$\mathbb{\tilde{\tilde{S}}}: \{h_T(x)\}\rightarrow \{\alpha(\lambda),\beta(\lambda) \},$$
by
$$
\left(
  \begin{array}{c}
    \beta(\lambda) \\
    \alpha(\lambda) \\
  \end{array}
\right)=\mu_1^{(2)}(0,T;\lambda)=\left(
  \begin{array}{c}
    \mu_1^{12}(0,T;\lambda) \\
    \mu_1^{22}(0,T;\lambda) \\
  \end{array}
\right),   Im\lambda\geq 0,
$$
where $\mu_1(x,T;\lambda)$ is the unique solution of the Volterra linear integral equation
$$\mu_1(x,T;\lambda)=I-\int_{0}^{x}e^{i\lambda^2(\xi-x)\hat{\sigma}_3}(V_1\mu_1)(\xi,T;\lambda)d\xi.$$
\end{de}

\begin{prop}
The spectral functions $\alpha(\lambda)$ and $\beta(\lambda)$ have the
following properties
\begin{description}
  \item[(i)] $\alpha(\lambda)$ and $\beta(\lambda)$ are analytic for $Im\lambda >0$ and continuous and bounded for $Im\lambda \geq 0$;
  \item[(ii)]$\alpha(\lambda)=1+O(\frac{1}{\lambda}), \beta(\lambda)=O(\frac{1}{\lambda})$ as $\lambda\rightarrow\infty$, $Im\lambda \geq 0$;
  \item[(iii)] $\alpha(\lambda)\overline{\alpha(\bar{\lambda})}-\beta(\lambda)\overline{\beta(\bar{\lambda})}=1$, $\lambda \in \mathbb{R}$;
\item[(iv)] $\alpha(-\lambda)=\alpha(\lambda),\beta(-\lambda)=-\beta(\lambda)$, $Im\lambda \geq 0$;
\item[(v)] $\mathbb{\tilde{\tilde{Q}}}=\mathbb{\tilde{\tilde{S}}}^{-1}: \{\alpha(\lambda),\beta(\lambda) \}\rightarrow \{h_T(x)\}$, the inverse map $\mathbb{\tilde{\tilde{S}}}$ of $\mathbb{\tilde{\tilde{Q}}}$ is defined by
\eqa
h_T(x)=2im_{T}(x)e^{2i\int_{0}^{x}\Delta_1(\xi,T)d\xi},\,
m_T(x)=\lim_{\lambda\rightarrow\infty}(\lambda M^{(T)}(x,\lambda))_{12},\eeqa
where $M^{(T)}(x,\lambda)$ is the unique solution of the following RHP (see \textbf{Remark 3.9}).
\end{description}
\end{prop}

\begin{rem}
Let
\eqa \begin{array}{l}
M_{+}^{(T)}(x,\lambda)=(\frac{\mu_{3}^{D_3\cup D_4}(x,\lambda)}{\alpha(\lambda)},\mu_{1}^{D_3\cup D_4}(x,\lambda)),\,Im\lambda \geq 0,\\
M_{-}^{(T)}(x,\lambda)=(\mu_{1}^{D_1\cup D_2}(x,\lambda),\frac{\mu_{3}^{D_1\cup D_2}(x,\lambda)}{\overline{\alpha(\bar{\lambda})}}), Im\lambda \leq 0.
\end{array}\label{slisp}\eeqa
$M^{(T)}(x,\lambda)$ is the unique solution of the following RHP
\end{rem}

\begin{itemize}
  \item $M^{(t)}(t,\lambda)=\left\{
                             \begin{array}{ll}
                              M_{-}^{(T)} (x,\lambda), & Im \lambda \geq0 \\
                              M_{+}^{(T)} (x,\lambda), & Im \lambda \leq0
                             \end{array}
                           \right.$ is a sectionally meromorphic function.
  \item $M_{+}^{(T)} (x,\lambda)=M_{-}^{(T)} (x,\lambda)J^{(T)} (x,\lambda)$, $\lambda\in \mathbb{R}$, and
\eqa
J^{(T)}(x,\lambda)=\left(\begin{array}{cc}
\frac{1}{\alpha(\lambda)\overline{\alpha(\bar{\lambda})}} & \frac{\beta(\lambda)}{\overline{\alpha(\bar{\lambda})}}e^{-2i(\lambda x+2\lambda^2T)} \\
-\frac{\overline{\beta(\bar{\lambda})}}{\alpha(\lambda)}e^{2i(\lambda x+2\lambda^2T)} & 1
\end{array}\right), \lambda \in \mathbb{R}.\label{slisp}\eeqa
  \item $M^{(T)} (x,\lambda)=I+O(\frac{1}{\lambda}), \lambda\rightarrow\infty.$
  \item $\alpha(\lambda)$ has $2N$ simple zeros $\{\gamma_j\}_{1}^{2N}$, $2N=2N_1+2N_2$, such that, $\gamma_j$($j=1,2,\cdots,2N_1$) lie in $D_1\cup D_2$), $\bar{\gamma}_j$($j=2N_1+1,2N_1+2,\cdots,2N$) lie in $D_3\cup D_4.$
  \item The first column of $M_{+}^{(T)}(x,\lambda)$ has simple poles at $\lambda=\gamma_j$, $j=1,2,\cdots,2N_1$. The second column of $M_{-}^{(T)}(x,\lambda)$ has simple poles at $\lambda=\bar{\gamma}_j$, $j=1,2,\cdots,2N_2$.\\
The associated residues are given by
\eqa  Res \{[M^{(T)}(x,\lambda)]_{1} ,  \gamma_j \}=\frac{e^{2i(\gamma_jx+2\gamma_j^2t)}}{\dot{\alpha}(\gamma_j)\beta(\gamma_j)}[M^{(T)}(x,\gamma_j)]_{2}, j=1,2,\cdots,2N_1,\eeqa
\eqa  Res \{[M^{(T)}(x,\lambda)]_{2} , \bar{\gamma}_j\}
=\frac{e^{-2i(\gamma_jx+2\gamma_j^2t)}}{\overline{\dot{\alpha}(\bar{\gamma}_j)}\overline{\beta(\bar{\gamma}_j)}}[M^{(T)}(x,\bar{\gamma}_j)]_{1}, j=1,2,\cdots,2N_2.\label{slisp}\eeqa
\end{itemize}

\subsection{The principal Riemann-Hilbert problem}

\begin{thm}
Let $u_0(x)\in \mathcal{S}(\mathbb{R^{+}})$ is a smooth function. Suppose that the function $g_0(t),g_1(t)$ are compatible with the function $u_0(t)$.
Define the spectral functions $S(\lambda)$ and $s(\lambda)$ in terms of $a(\lambda)$, $b(\lambda)$, $A(\lambda)$ and $B(\lambda)$ in Eq.(2.31),
and define the spectral functions $a(\lambda)$, $b(\lambda)$, $A(\lambda)$ and $B(\lambda)$, in terms of $u_0(x)$ $g_0(t)$, $g_1(t)$ of \textbf{Defintion 3.1} and \textbf{Defintion 3.3}. We have the following global relation
$$ S^{-1}(\lambda)s(\lambda)=e^{2i\lambda^2T\hat\sigma_3}c(T,\lambda),\quad\lambda\in(D_2\cup D_4,D_1\cup D_3),$$
where $c(T,\lambda)=\mu_3(0,t,\lambda)$ and $s(\lambda)=\mu_3(0,0;\lambda),S(\lambda)=S(T,\lambda)=(e^{2i\lambda^2T\hat\sigma_3}\mu_2(0,T;\lambda))^{-1}$ defined in (2.27), if $\lambda \rightarrow \infty$ the global relation is replace by $a(\lambda)B(\lambda)-b(\lambda)A(\lambda)=0$.
Assume that the possible zeros $\{\varepsilon_j\}_{j=1}^{2n}$ of $a(\lambda)$ and $\{\gamma_j\}_{j=1}^{2N}$ of $\alpha(\lambda)$, define the $M(x,t,\lambda)$ as the solution of the following Riemann-Hilbert problem
\begin{itemize}
  \item $M(x,t;\lambda)$ is a sectionally meromorphic on the Riemann $\lambda$-sphere with jumps across
the contours on $\bar D_n\cap\bar D_m(n,m=1,2,3,4)$ (see figure 2).
  \item $M(x,t;\lambda)$ satisfies the jump condition with jumps across the contours on $\bar D_n\cap\bar D_m(n,m=1,2,3,4)$
\beq M_{+}(x,t;\lambda)=M_{-}(x,t;\lambda)J(x,t;\lambda),\,\lambda \in \bar D_n\cap\bar D_m,\,n,m=1,2,3,4; n\neq m.\label{slisp}\eeq
\item $M(x,t;\lambda)=I+O(\frac{1}{\lambda}),\,\lambda\rightarrow\infty$.
\item The residue condition of $M(x,t;\lambda)$ is showed in \textbf{Proposition 2.6}.
\end{itemize}
Then $M(x,t;\lambda)$ exists and is unique, we define $u(x,t)$ in terms of $M(x,t;\lambda)$ by
\eqa \begin{array}{l}
u(x,t)=2im(x,t)e^{2i\int_{(0,0)}^{(x,t)}\Delta},\\
m(x,t)=\lim_{\lambda\rightarrow\infty}(\lambda M(x,t;\lambda))_{12},\\
\Delta=4|m|^2dx+(48|m|^4-4i(\bar{m}_xm-m_x\bar{m}))dt.
\end{array}\label{slisp}\eeqa
Furthermore, under the defined that the initial values $u(x,0)=u_0(x)$ and boundary values  $u(0,t)=g_0(t)$, $u_x(0,t)=g_1(t)$ lie in the Schwartz space, $u(x,t)$ is the solution of the KE equation (1.1).
\end{thm}

\textbf{Proof.} In fact, if $a(\lambda)$ and $\alpha(\lambda)$ have no zeroes, then the $2\times2$ function $M(x,t;\lambda)$ satisfies a non-sigular RHP. Using the fact that the jump matrix $J(x,t;\lambda)$
match with the symmetry conditions, we can show that this problem has a unique global solution \cite{Adler1997}. The case that $a(\lambda)$ and $\alpha(\lambda)$ have a finite number of zeros can be mapped to the case of no zeros supplemented by an algebraic system of equations which is always uniquely solvable.

Therefore, similar to the reference \cite{Zhang2017}, we have the following vanishing theorem hold true.

\begin{thm}
 The RHP in \textbf{Theorem 3.10} with the vanishing boundary condition $M(x,t;\lambda)\rightarrow 0$ ($\lambda \rightarrow \infty)$, has only the zero solution.\\
\textbf{Proof} Suppose that $M(x,t;\lambda)$ is a solution of the RHP in \textbf{Theorem 3.10} such that $M_{\pm}(x,t;\lambda)\rightarrow\infty$ ($\lambda \rightarrow \infty)$. Let $A$ is a $2\times2$ matrix, $A^{\dag}$denote
the complex conjugate transpose of $A$. Define
\begin{equation}\begin{array}{cc}
H_{+}(\lambda)=M_+(\lambda)M_{-}^{\dag}(-\bar{\lambda}),\,Im\lambda^2\geq0,\\
H_{-}(\lambda)=M_-(\lambda)M_{+}^{\dag}(-\bar{\lambda}),\,Im\lambda^2\leq0,
\end{array}\end{equation}
where the $x$ and $t$ are dependence. $H_{+}(\lambda)$ and $H_{-}(\lambda)$ is analytic in $\{\lambda\in \mathbb{C} \setminus Im\lambda^2>0\}$ and $\{\lambda\in \mathbb{C} \setminus Im\lambda^2<0\}$,
respectively. By the symmetry relations $a(-\lambda)=a(\lambda),b(-\lambda)=-b(\lambda),$ and $A(-\lambda)=A(\lambda),B(-\lambda)=-B(\lambda)$ in \textbf{Proposition 2.3} and Eq.(2.36), we have
\begin{equation}
J_{1}^{\dag}(-\bar{\lambda})=J_1(\lambda),\,J_{3}^{\dag}(-\bar{\lambda})=J_3(\lambda),\,J_{2}^{\dag}(-\bar{\lambda})=J_4(\lambda).
\end{equation}
Then
\begin{equation}
\begin{array}{cc}
H_{+}(\lambda)=M_-(\lambda)J(\lambda)M_{-}^{\dag}(-\bar{\lambda}),\,Im\lambda^2\in \mathbb{R},\\
H_{-}(\lambda)=M_-(\lambda)J^{\dag}(-\bar{\lambda})M_{-}^{\dag}(-\bar{\lambda}),\,,Im\lambda^2\in \mathbb{R}.
\end{array}
\end{equation}
It is not difficult to see that $H_{+}(\lambda)=H_{-}(\lambda)$ for $Im\lambda^2\in \mathbb{R}$. Therefore, $H_{+}(\lambda)$ and $H_{-}(\lambda)$ define an entire function vanishing at infinity, so $H_{+}(\lambda)$ and $H_{-}(\lambda)$ are identically zero. Noting $J_3(i\kappa)(\kappa\in\mathbb{R}),$ is a Hermitian matrix with unit determinant and $(2,2)$ entry $1$ for any $\kappa\in\mathbb{R}$. Therefore, $J_3(i\kappa)(\kappa\in\mathbb{R})$
is a positive definite matrix. Since $H_{-}(\kappa)$ vanishes identically
for $\kappa\in i\mathbb{R}$, i.e.
\begin{equation}
M_+(i\kappa)J_3(i\kappa)M_{+}^{\dag}(i\kappa)=0,\,\kappa\in\mathbb{R}.
\end{equation}
We can infer that $M_+(i\kappa)=0$ as $\kappa\in\mathbb{R}$. It follows that $M_{+}(\lambda)$ and $M_-(\lambda)$ vanish identically.
\end{thm}

\begin{rem} Proof that $u(x,t)$ satisfies the KE equation.\end{rem}

Using arguments of the dressing method \cite{Zakharov1979}, it can be verified directly that if $u(x,t)$ is defined in terms of $M(x,t;\lambda)$ by Eq.(3.17), and if $M(x,t;\lambda)$ is defined as the unique solution of the above RHP, then $u(x,t)$ and $M(x,t;\lambda)$ satisfy two parts of the Lax pair, hence $u(x,t)$ is solvable on KE equation.

\begin{rem} We can use similar method with \cite{xu2014} to prove that initial values $ u(x,0)=u_0(x)$ and boundary values $u(0,t)=g_0(t), u_x(0,t)=g_1(t)$, we omit this proof in here because of the length of this article.
\end{rem}

\section{Conclusions and discussions}

In this paper, we consider IBV of the KE equation on the half-line. Using the Fokas unified transform method for nonlinear evolution systems which taking the form of Lax pair isospectral deformations and whose corresponding continuous spectra Lax operators, assume that the solution $u(x,t)$ exists, we show that it can be represented in terms of the solution of a matrix Riemann-Hilbert problem formulated in the plane of the complex spectral parameter $\lambda$. The jump matrix has explicit $(x,t)$ dependence and is given in terms of the spectral functions $a(\lambda),b(\lambda)$ and $A(\lambda),B(\lambda)$, which obtained from the initial data $u_0(x)=u(x,0)$ and the boundary data $g_0(t)=u(0,t),g_1(t)=u_x(0,t)$, respectively. The spectral functions are not independent, but related by a compatibility condition, the so-called global relation.
For other integrable equations, can we construct their solution of a matrix Riemann-Hilbert problem formulated in the plane of the complex spectral parameter $\lambda$ by the similar method?
In paper \cite{Xiao2016}, Xiao and Fan obtained the soliton-type solutions of the Harry-Dym equation by solving the particular Riemann-Hilbert problem with vanishing scattering coefficients, can we obtain the soliton-type solutions of the KE equation following the same ways?
Moreover, Zhu have successfully applied the Deift-Zhou nonlinear steepest descent method to analyzed the long-time asymptotic for the solutions of decay initial value problem of the KE equation in \cite{Zhu2018}. But under the assumption that the initial and boundary values lie in the Schwartz class, can we do the long-time asymptotics for the solutions of the decay initial and boundary values of KE equation following the same ways as for the derivative nonlinear Schr\"{o}dinger equation \cite{Arruda2017} and Fokas-Lenells equation \cite{yan2017}? These three questions will be discussed in our future work.

\subsection*{Acknowledgements}

This work is partially supported by the National Natural Science Foundation of China under
Grant Nos. 12271008 and 11601055, Natural Science Foundation of Anhui Province under
Grant No.1408085QA06.

\end{document}